\documentclass{bmvc2k}
\usepackage{multirow}
\usepackage{todonotes}
\usepackage{arydshln}

\usepackage{amssymb}
\title{Self-Supervised Curricular Deep Learning for Chest X-Ray Image Classification}

\addauthor{Iván de Andrés Tamé}{ivan.andrest@estudiante.uam.es}{1}
\addauthor{Kirill Sirotkin}{kirill.sirotkin@uam.es}{1}
\addauthor{Pablo Carballeira}{pablo.carballeira@uam.es}{1}
\addauthor{Marcos Escudero-Viñolo}{marcos.escudero@uam.es}{1}
\addinstitution{
 Video Processing and Understanding Lab\\
 Escuela Politécnica Superior\\
 Universidad Autónoma de Madrid
 \\28049 Madrid, Spain
}

\runninghead{de Andrés, et al.}{SSL Curricular DL for CXR Image Classification}

\def\etal{\emph{et al}\bmvaOneDot}

\begin{document}

\maketitle

\begin{abstract}
Deep learning technologies have already demonstrated a high potential to build diagnosis support systems from medical imaging data, such as Chest X-Ray images. However, the shortage of labeled data in the medical field represents one key obstacle to narrow down the performance gap with respect to applications in other image domains. In this work, we investigate the benefits of a curricular Self-Supervised Learning (SSL) pretraining scheme with respect to fully-supervised training regimes for pneumonia recognition on Chest X-Ray images of Covid-19 patients. We show that curricular SSL pretraining, which leverages unlabeled data, outperforms models trained from scratch, or pretrained on ImageNet, indicating the potential of performance gains by SSL pretraining on massive unlabeled datasets. Finally, we demonstrate that top-performing SSL-pretrained models show a higher degree of attention in the lung regions, embodying models that may be more robust to possible external confounding factors in the training datasets, identified by previous works~\cite{afifi2021ensemble, arias2021covid}.
\end{abstract}

\section{Introduction} \label{section:intro}

The medical imaging field is constantly evolving, and solutions for diagnosis support systems, based on deep learning technologies bear many promises. However, even though the amount of collected medical data has never been larger, it is still (i) much smaller than in other application domains, (ii) often not publicly available and (iii) often not annotated. All these issues together impose significant constraints on the use of supervised deep learning solutions.
Specifically, application scenarios for Chest X-Ray images (CXR)  include \cite{ccalli2021deep}: classification and localization of pathologies, image segmentation to split the relevant region of the image to diagnose, and the generation of new synthetic images for data augmentation, or resolution increase.
% \todo[inline,color=pink]{pcl:  yo lo pondria como el ejemplo que te he puesto. Asi te va a comer mucho espacio\\ iat:asi o con salto de linea?? Porque asi parece que se pone en negrita solo}
\paragraph{Self-Supervised Learning}
Generally, the performance of supervised deep learning methods is directly related to the size of the labeled training dataset. This may harm its performance on CXR data, as the annotation of this data modality is a heavy burden that requires scarce and expensive expertise. Therefore, CXR data represent a paradigmatic modality for a current trend to improve the performance of visual tasks with a shortage of labeled data: Self-Supervised Learning (SSL) techniques. The hypothesis is that a combination of SSL pretraining on unlabeled data, followed by a fine-tuning step over the same target data can improve the performance of a model~\cite{jaiswal2020survey}, as the SSL training aids to learn meaningful representations that act as an advantageous starting point for learning the target task.

For example in~\cite{doersch2015unsupervised}, it is claimed that by predicting relative locations of randomly sampled image patches the model learns high-level image representations. More recent works propose to learn efficient representations by matching the augmentations of the same image using the contrastive loss function. To avoid trivial solutions, SimCLR~\cite{chen2020simple} uses very large batches during training increasing the number of negative samples. Momentum Contrast (MoCo)~\cite{he2020momentum} introduces a memory queue that allows to decrease the batch size. Finally, BYOL~\cite{grill2020bootstrap} and SimSiam~\cite{chen2021exploring} rely only on positive pairs and propose other ways of preventing degenerate solutions, such as a momentum encoder and stop-gradient operations. Recent advances in SSL are closing the performance gap with the purely supervised models in natural imagining domains~\cite{chen2021empirical, caron2021emerging, chen2020big}. Moreover, SwAV~\cite{caron2020unsupervised} demonstrated that self-supervised features learned from ImageNet can outperform the supervised ones in transfer learning setups with Places205~\cite{zhou2014learning}, PASCAL VOC~\cite{everingham2010pascal} and iNaturalist2018~\cite{van2018inaturalist} datasets. SSL pretraining has been assessed in several different medical domains. For instance, image-level diagnosis on CXR images has been explored in several works~\cite{gazda2021self, park2021deep}. Another study~\cite{kwasigroch2020self} proposed to use Rotation Prediction~\cite{gidaris2018unsupervised} and Jigsaw puzzles~\cite{noroozi2016unsupervised} to aid melanoma recognition on ISIC-2017 dataset, and more a recent work~\cite{matsoukas2021time} investigates applications of DINO~\cite{caron2021emerging} to three standard medical imaging datasets: APTOS-2019~\cite{kaggle}, ISIC-2019~\cite{codella2018skin} and CBIS-DDSM~\cite{lee2017curated}.

% \todo[inline,color=yellow]{This paragraph is proposed by Kirill but I dont know if is good idea to talk that much of skin solutions here. May this can be removed if we need space}
% \todo[inline,color=green]{pcl: I like the text, but we probably need to reduce it. Finish formatting it, and then we'll see what we need to reuce}

Some works have proposed to combine different pretext tasks to improve the performance of single-task SSL pretraining~\cite{doersch2017multi, sirotkin2021improved, liu2019end, zhu2020aggregative}. The intuition behind combining pretext tasks is that different pretext tasks learn different visual representations from the same images~\cite{wang2020self}, and thus, these different features contribute to a better representation of the data, and a better starting point for a downstream task. Doersch and Zisserman~\cite{doersch2017multi} combine four different SSL tasks at a batch level, i.e, each SSL task trains over the same batch before updating the network weights using a combination of the gradients. This strategy shows a performance improvement with respect to single-task pretraining on three datasets: ImageNet~\cite{krizhevsky2012imagenet}, PASCAL VOC~\cite{everingham2010pascal}, and NYU depth prediction~\cite{silberman2012indoor}. In~\cite{sirotkin2021improved}, Sirotkin \etal. propose a curricular SSL pretraining method, where a CNN backbone is trained sequentially, using multiple SSL models. One of the main differences with respect to~\cite{doersch2017multi} is that this approach involves training on the complete dataset before changing the SSL task. This method is tested on the ISIC-2017 binary and ISIC-2019 multi-class skin lesion classification challenges~\cite{tschandl2018ham10000, codella2018skin}. The authors provide evidence of performance improvement with respect to single-task pretraining, which depends on the ordering of pretraining pretext tasks.

In this work we explore the benefits of using a curricular SSL training scheme for pneumonia classification on CXR images of Covid-19 patients. We demonstrate that the best curricular-SSL pretraining achieves a balanced accuracy of 85.67\%, which outperforms pretraining the model on ImageNet (82.75\%) and random model initialization (83.69\%).
% \todo[inline,color=green]{pcl: re-write the following part: dont say we follow an ordering, say that this best combination corresponds to a curriculum ordering of tasks, as defined in REF}
% The curricular ordering is defined in~\cite{sirotkin2021improved}, and consists in ordering the SSL tasks according to their individual accuracy. 

\paragraph{Covid-19 pneumonia in CXR data}
Predominant findings of pneumonia in CXRs of Covid-19 patients are opacities inside the lung regions, which are defined as ground-glass opacities (GGO)~\cite{litmanovich2020review, wong2020frequency}. Empirical evidence~\cite{bhattacharya2022radiotransformer} shows that radiologists themselves diagnose mainly based on the CXR data inside the lung regions. However, there are indications that recent deep learning systems targeting disease detection from CXRs, rather than learning  on  the medical pathology evidence, rely on confounding factors~\cite{degrave2021ai}, out of the lung regions, as a learning shortcut. The confounding factors are prone to be dependent on the training dataset, so to avoid models that are overfitted to a given training dataset, and to increase the scalability and generalization capabilities of these solutions, there is a need to develop models that are robust to such confounding factors. One approach to indicate that a model is less affected by these external factors is the amount of attention it focuses on lung area. 
%in which the GGOs are, i.e. lungs. 

In this work we propose a novel strategy to quantitatively compare different models in terms of the degree of attention they present in the lung regions. This strategy is used to show evidence that SSL pretraining (and currcular SSL pretraining) is beneficial to focus the model's attention on the region of interest of the CXR image, in this case, the lungs. Therefore, SSL-pretrained models are prone to be more robust to the aforementioned external confounding factors. 

\section{Experimental design} \label{section:methodology}

\subsection{Curricular SSL training scheme}

\begin{figure}[]
    \centering
    \includegraphics[width=\linewidth]{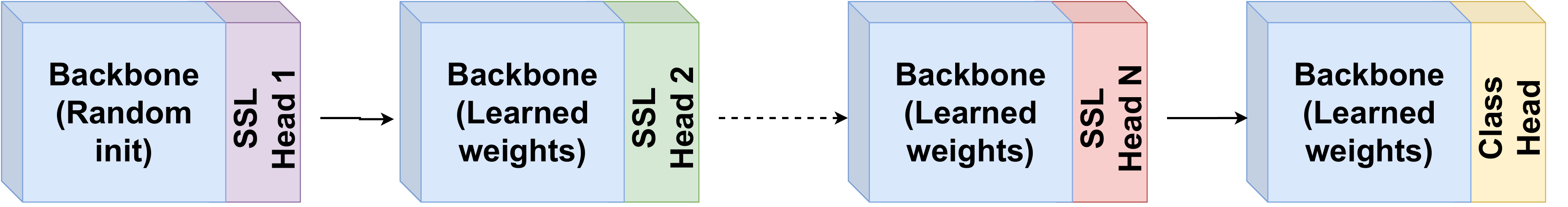}
    \caption[Training scheme proposed by Sirotkin et al.]{Curricular SSL training scheme, proposed in~\cite{sirotkin2021improved}. It uses a sequence of SSL tasks as a pretraining stage, and a final fine-tuning classification stage (downstream task).}
    \label{model}
\end{figure}

We have used the curricular SSL training scheme proposed in~\cite{sirotkin2021improved} and depicted in Figure~\ref{model}. The training pipeline is composed of $N$ pretraining steps, each being an SSL task, and a final training on the downstream task, which in our work is a classification task. For all of the training steps we use the same backbone architecture, along with the same dataset. After each pretraining step, the backbone weights are transferred to the next step, making the model combine the learning from each SSL task to support the final classification. Training hyperparameters, e.g. learning rate or batch size, have a direct influence on the training pace and the final performance of a model. In a curricular SSL training, such as the one depicted on Figure~\ref{model}, the hyper-parameters of each step may influence the subsequent ones. In this work, we have selected the learning-rate (LR) of each training step (both SSL steps and downstream classification) using the following policy:
\begin{itemize}
    \item Each training step is repeated, using different LR values taken from a pre-defined range, for a limited number of epochs.
    \item The LR that leads to the highest performance (or lowest loss value) on the training task is used to train the model for the full number of epochs.  
\end{itemize}

\subsection{Comparative analysis of the attention of the models: AIL Score} \label{subsection:attention}

\begin{figure}[]
    \centering
    \includegraphics[width=\linewidth]{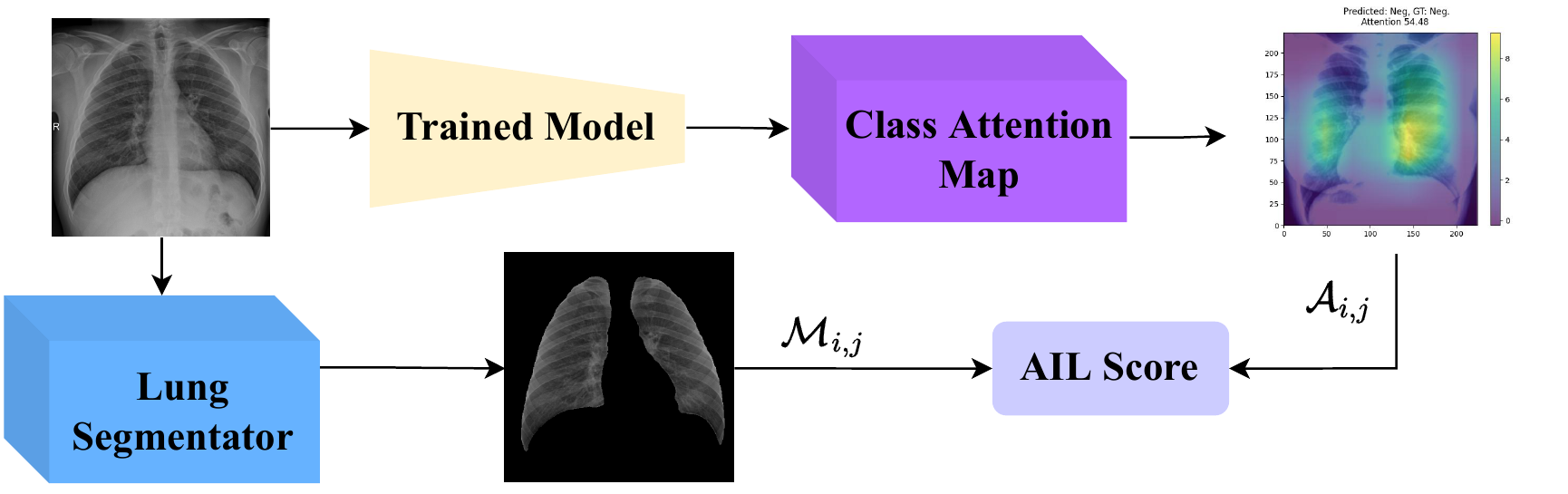}
    \caption{Overview of the processing scheme used to compute the AIL score. Note that, although we have included a lung texturized mask image (it allows to verify the correct lung segmentation), we use the binary segmentation mask computation of the AIL score.}
    \label{model_overview}
\end{figure}

As detailed in Section~\ref{section:intro}, deep learning systems for pathology detection in CXRs may rely on confounding factors~\cite{degrave2021ai}. Moreover, other independent works ~\cite{afifi2021ensemble, arias2021covid} demonstrate the same trend in the BIMCV-COVID19+ dataset\textemdash a part of the dataset used in this work (see details in Section~\ref{section:results}). For the sake of scalability and utility, it is desirable to build models that focus their attention on the lung regions, ignoring the existence of external confounding factors in the training dataset. This presents an important advantage when the distribution of validation data differs from that of the training data. Specifically, if a model performs the classification prediction based on the data in the lung regions, it is prone to be more robust to background changes produced, for instance, by a change in the setup of a capturing device.

In this work, we investigate the degree of attention in the lung regions presented by models trained with different cofigurations of curricular SSL training schemes and fully-supervised ones. This is done through an attention score that measures the cumulative attention of a model with respect to the cumulative attention in the whole image: \textit{Attention Inside the Lungs} (AIL, Equation~\ref{equation_AIL}). Figure~\ref{model_overview} shows the processing scheme use to compute this AIL score. First, a Class Attention Map (CAM)~\cite{zhou2016learning} (predicted class), $\mathcal{A}_{i,j}$, is obtained from an input image and a model. Second, a binary mask  of the lung regions, $\mathcal{M}_{i,j}$, is obtained using a state-of-the-art lung segmentation model. The AIL score is obtained as follows: 

\begin{equation}
    AIL = \frac{ \sum_{i,j} \mathcal{A}_{i,j}\times \mathcal{M}_{i,j}}{\sum_{i,j} \mathcal{A}_{i,j}}.
    \label{equation_AIL}
\end{equation}

The AIL score allow us to compare the level attention of in-the-lung regions of several models. Higher AIL values correspond to CAMs with a higher focus in the lung regions. 

% \begin{equation}
%     AIL = \frac{ \sum_{i,j} I_{mask}(CAM^{img}_{i,j}) }{\sum_{i,j} CAM^{img}_{i,j}}
%     \label{equation_AIL}
% \end{equation}

\begin{figure}[]
    \centering
    \begin{tabular}{ccc}
    \includegraphics[width=0.2\linewidth]{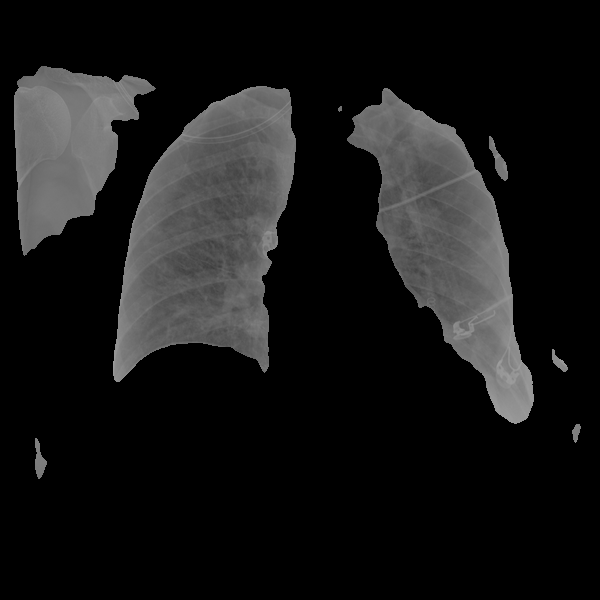} &
    \includegraphics[width=0.2\linewidth]{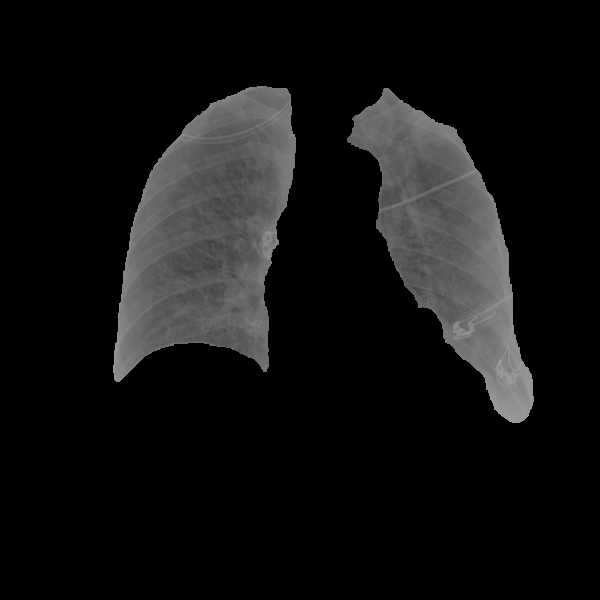} &
    \includegraphics[width=0.2\linewidth]{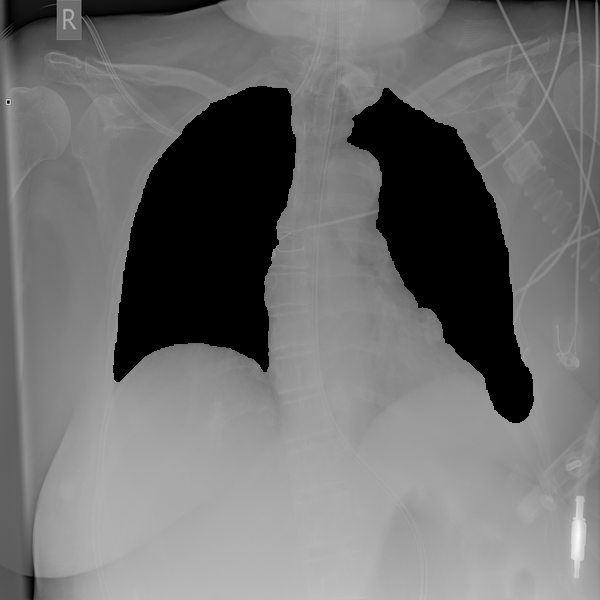} \\
    (a) & (b) & (c)\\
\end{tabular}
    \caption{Examples of lung mask obtained using~\cite{selvan2020lung} before (a) and after (b) the post-processing step. (c) Example of an inversely-segmented image used, in Section \ref{subsection:ail}, to train a model excluding the lung region data (evidence of external confoundig factors).}
    \label{masks}
\end{figure}

In our implementation, lung segmentation has been carried out using the solution proposed by Selvan et al.~\cite{selvan2020lung}, based on an U-net architecture and a variational encoder for data imputation, trained on public datasets labeled for tuberculosis detection~\cite{jaeger2014two}, and specifically the model provided by the authors\footnote{\url{https://github.com/raghavian/lungVAE/tree/master/saved_models}}. The segmentation masks are post-processed using connected component analysis to exclude small erroneous regions and binary morphological closing to ﬁll any small holes in the masks (see Figure~\ref{masks} for examples).

\section{Experimental results} \label{section:results}

\subsection{Dataset and performance metrics} \label{subsection:dataset}

%\todo[inline,color=green]{pcl: Add the dataset \% next to each label}

We have used the SIIM-FISABIO-RSNA COVID-19 Detection dataset~\cite{lakhani20212021}, which is provided by the Society for Imaging Informatics in Medicine (SIIM) and collects CXR images of Covid-19 patients. Images are labeled for the identification and location of pneumonia abnormalities, which are classified into four groups\footnote{A detailed explanation of each label can be found in \cite{lakhani20212021}}: \textit{Negative for pneumonia} (28\%) images have no lung opacities, \textit{Typical appearance} (47\%) are images with presence of bilateral opacities, \textit{Indeterminate appearance} (17\%) have unilateral or central opacities, and \textit{Atypical appearance} (8\%) are different diseases without features of pneumonia. The SIIM-FISABIO-RSNA training data consists of  6,334 chest scans and is built from two datasets: BIMCV-COVID19+~\cite{vaya2020bimcv} and MIDRC-RICORD~\cite{tsai2021rsna}.

Original images are in the DICOM format (Digital Imaging and Communication In Medicine) with a resolution over 2K. In our experiments, images have been cropped and resized to a 224$\times$224 resolution, and an 80\%-20\% train-val split has been used. Performance is assessed in terms of balanced accuracy.

% \begin{table}[]
% \centering
% \begin{tabular}{|l|l|c|}
% \hline
% \textbf{Label} & \textbf{Description} & \textbf{\%} \\ \hline
% \begin{tabular}[c]{@{}l@{}}Negative for\\ pneumonia\end{tabular} & No lung opacities & 28\% \\ \hline
% \begin{tabular}[c]{@{}l@{}}Typical \\ appearance\end{tabular} & \begin{tabular}[c]{@{}l@{}}Bilateral, peripheral, multi-focal predominant opacities\\ Diffuse bilateral opacities including both central and peripheral.\\ Diffuse bilateral opacities with fibrosis / reduced lung volumes\end{tabular} & 47\% \\ \hline
% \begin{tabular}[c]{@{}l@{}}Indeterminate\\ appearance\end{tabular} & \begin{tabular}[c]{@{}l@{}}Upper lung zone predominant opacities \\ Unilateral opacities, even if multi-focal\\ Central opacities with relative peripheral sparing\end{tabular} & 17\% \\ \hline
% \begin{tabular}[c]{@{}l@{}}Atypical\\ appearance\end{tabular} & \begin{tabular}[c]{@{}l@{}}Pneumothorax without features of pneumonia.\\ Pleural effusion without features of pneumonia.\\ Mass(es) or nodule(s) \\ Lobar Pneumonia\\ Scarring / fibrosis\end{tabular} & 8\% \\ \hline
% \end{tabular}
% \caption{Labels from the SIIM-FISABIO-RSNA COVID-19 dataset, their meanings, and the \% of appearance in the dataset}

% \label{labels}
% \end{table}

\subsection{Curricular SSL training scheme configuration}

In our experiments, we have used ResNet-50~\cite{he2016deep} architecture for the backbone of the model trained with the scheme described in Section~\ref{section:methodology}. This architecture has been used in several recent works~\cite{showkat2022efficacy, shelke2021chest} on pneumonia recognition in Covid-19 patients, and shows competitive performance with respect to DenseNet-121~\cite{huang2017densely}\textemdash another common choice for the task~\cite{ridzuan2022challenges}. Besides, ResNet-50 is less computational-demanding than DenseNet-121, which allows to perform exhaustive training experiments, as those required for measuring the effectiveness and biases of different combinations of SSL tasks and LRs, as the main objective of the experiments is to evaluate the performance gain obtained through curricular SSL pretraining.

Four different SSL pretext tasks, and some of their combinations have been explored, namely: MoCo v2~\cite{chen2020improved}, SwAV~\cite{caron2020unsupervised}, Relative Location~\cite{doersch2015unsupervised} and Rotation Prediction~\cite{gidaris2018unsupervised}, using the MMSelfSup framework~\cite{mmselfsup2021}. Table~\ref{hyperparams} collects the training hyperparameters that are used in each training step. A weighted cross-entropy loss, following class ratios detailed in Subsection \ref{subsection:dataset}, guides the training process. For every training scheme, and training step, we perform the LR search described in Section \ref{section:methodology}. The number of epochs for the LR search has been selected, per task, as the minimum number of epochs required to establish an ordering based on performance (which is maintained afterwards). The number of epochs for full-training is selected to ensure training convergence.

% to know which LR will be top performance when training the complete number of epochs.

\begin{table}[]
\centering
\resizebox{\textwidth}{!}{%
\begin{tabular}{cccccccc}
\hline
Task & \begin{tabular}[c]{@{}c@{}}Batch\\ Size\end{tabular} & \begin{tabular}[c]{@{}c@{}}LR-Search \\ epochs\end{tabular} & LR Range & \begin{tabular}[c]{@{}c@{}}Full-training \\ epochs\end{tabular} & Optimizer & Momentum & \begin{tabular}[c]{@{}c@{}}Weight\\ decay\end{tabular} \\ \hline
Classification & 64 & 80 & {[}0.01-0.25{]} & 150 & SGD & 0.9 & 1e-4 \\ \hline
Relative Location & 32 & 20 & {[}0.01-0.25{]} & 30 & SGD & 0.9 & 1e-4 \\ \hline
Rotation Prediction & 16 & 20 & {[}0.01-0.25{]} & 30 & SGD & 0.9 & 1e-4 \\ \hline
MoCo v2 & 32 & 20 & {[}0.01-0.25{]} & 30 & SGD & 0.9 & 1e-4 \\ \hline
SwAV & 8 & 20 & {[}0.01-0.25{]} & 30 & LARS & 0.9 & 1e-4 \\ \hline
\end{tabular}%
}
\caption{Values of training hyperparameters for the training tasks. Fixed for all experiments.
}
\label{hyperparams}
\end{table}

As a reference, we compare the SSL-pretrained results with two different baselines: training the classifier from initial random weights (\textit{Scratch}), and training it from  \textit{ImageNet} weights (following the same LR search procedure).

% We selected a different number of epochs for the model to train and for the early stopping depending on the task that it is solving. 

% We set this number as the 
% After the early-stopping we keep the best LR and train for the complete number of epochs.

\subsection{Discussion on the labeling consistency of the SIIM-FISABIO-RSNA COVID-19 Detection dataset }\label{section:sound}

\begin{figure}
\centering
\begin{tabular}{ccc}
    \includegraphics[width=0.2\linewidth]{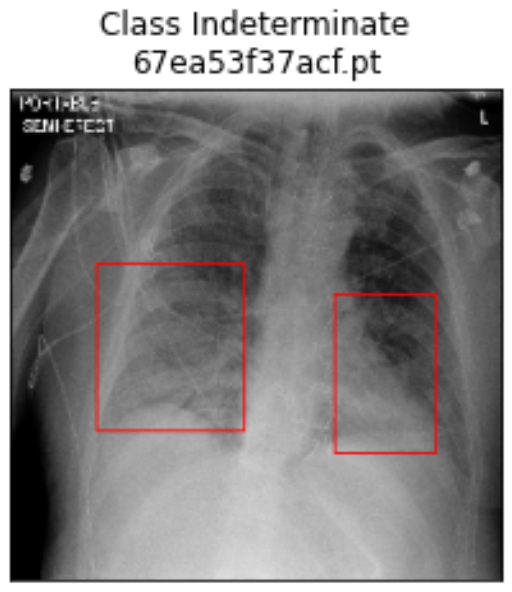} &
    \includegraphics[width=0.2\linewidth]{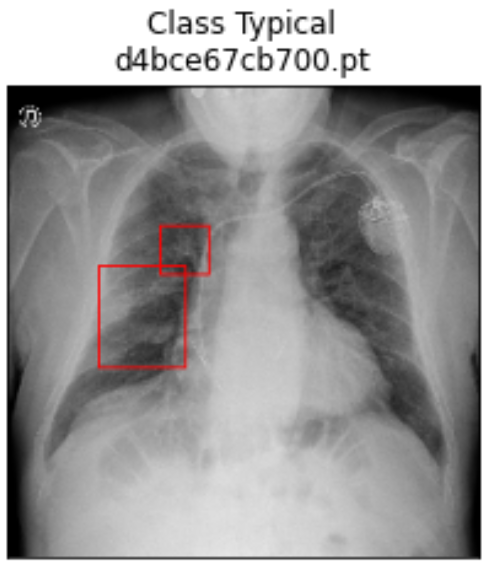} & \includegraphics[width=0.2\linewidth]{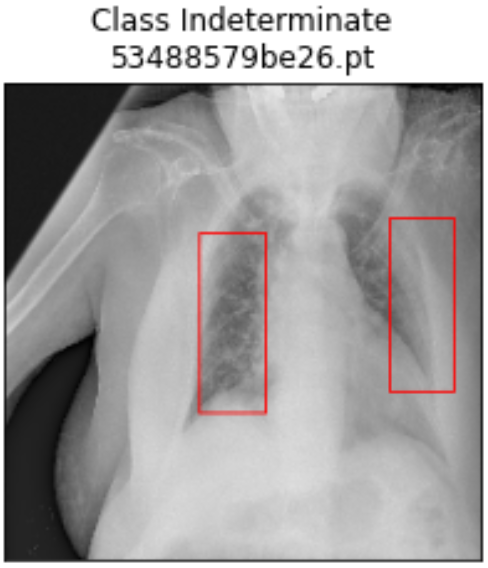} \\
    (a) & (b) & (c)\\
\end{tabular}
\caption{Examples of labeling inconsistencies taken from~\cite{ridzuan2022challenges}: (a) bilateral opacity (nominally \textit{Typical}) labeled as \textit{Indeterminate}; (b) unilateral opacity (nominally \textit{Indeterminate}) labeled as \textit{Typical}; (c) bilateral, lower-to-central lung opacity (nominally \textit{Typical}) labeled as \textit{Indeterminate}.} 
\label{bad_annotation}
\end{figure}

Existing literature indicates that the SIIM-FISABIO-RSNA COVID-19 Detection dataset may present labeling inconsistencies regarding the positions of the opacities~\cite{ridzuan2022challenges}. In particular, the annotations often confuse the \textit{Typical} and \textit{Indeterminate} classes, as they both correspond to pneumonia pathologies: i.e, the images labeled as \textit{Typical} have the common symptoms of pneumonia (bilateral opacities), while images labeled as \textit{Indeterminate} correspond to pneumonia but are characterized by uncommon symptoms. Examples of labeling inconsistencies are shown in Figure~\ref{bad_annotation}. This inconsistency may be particularly problematic for the \textit{Indeterminate} class, as the number of images (1076) is lower than the \textit{Typical} class~(2977). Besides, the \textit{Atypical} class collects all pathological images that do not correspond to pneumonia, thus it may present a high intra-class variation. This, coupled with the shortage of training data belonging to the \textit{Atypical} class (506 images) presents an impediment for an effective learning of class features by the model.

Considering the aforementioned issues, we first evaluate the capability of the CNN architecture to learn effective representations for all classes, and thus to achieve recognition rates beyond random-prediction on all of them. We train the supervised baselines and compare their performance with the SSL-pretrained models, as shown in Table~\ref{training}. The results indicate that the model is unable to fit the \textit{Indeterminate appearance} and \textit{Atypical appearance} classes, not even for training data, yielding lower than random performance. Hereinafter, we restrict our training and evaluation processes to the training and classification of the two majority classes (\textit{Negative} and \textit{Typical}). Thereby, we avoid interferences caused by inconsistently-labeled problems that can lead to results missinterpretations and may occlude the hypothetical benefits of using SSL.

\begin{table}[]
\centering
\resizebox{\textwidth}{!}{
\begin{tabular}{|c|cc|cc|cc|cc|cc|}
\hline
\multirow{2}{*}{\textbf{Pretraining}} & \multicolumn{2}{c|}{\textbf{\begin{tabular}[c]{@{}c@{}}Negative\\ Acc, \%\end{tabular}}} & \multicolumn{2}{c|}{\textbf{\begin{tabular}[c]{@{}c@{}}Typical\\ Acc, \%\end{tabular}}} & \multicolumn{2}{c|}{\textbf{\begin{tabular}[c]{@{}c@{}}Indeterminate\\ Acc, \%\end{tabular}}} & \multicolumn{2}{c|}{\textbf{\begin{tabular}[c]{@{}c@{}}Atypical\\ Acc, \%\end{tabular}}} & \multicolumn{2}{c|}{\textbf{\begin{tabular}[c]{@{}c@{}}Mean\\ Acc, \%\end{tabular}}} \\ \cline{2-11} 
 & \multicolumn{1}{c|}{\textbf{Train}} & \textbf{Validation} & \multicolumn{1}{c|}{\textbf{Train}} & \textbf{Validation} & \multicolumn{1}{c|}{\textbf{Train}} & \textbf{Validation} & \multicolumn{1}{c|}{\textbf{Train}} & \textbf{Validation} & \multicolumn{1}{c|}{\textbf{Train}} & \textbf{Validation} \\ \hline
\textbf{Scratch} & \multicolumn{1}{c|}{85.82} & 74.36 & \multicolumn{1}{c|}{82.02} & 67.86 & \multicolumn{1}{c|}{\textbf{19.60}} & \textbf{0.46} & \multicolumn{1}{c|}{\textbf{25.33}} & \textbf{0.99} & \multicolumn{1}{c|}{53.19} & 35.923 \\ \hline
\textbf{ImageNet} & \multicolumn{1}{c|}{84.24} & 79.74 & \multicolumn{1}{c|}{85.26} & 82.13 & \multicolumn{1}{c|}{\textbf{15.90}} & \textbf{6.07} & \multicolumn{1}{c|}{\textbf{29.50}} &\textbf{0.00} & \multicolumn{1}{c|}{53.72} & 41.988 \\ \hline
\textbf{RelativeLoc} & \multicolumn{1}{c|}{85.86} & 81.32 & \multicolumn{1}{c|}{83.57} & 84.75 & \multicolumn{1}{c|}{\textbf{17.90}} & \textbf{0.00} & \multicolumn{1}{c|}{\textbf{47.80}} & \textbf{0.00} & \multicolumn{1}{c|}{58.78} & 41.521 \\ \hline
\textbf{RotationPred} & \multicolumn{1}{c|}{80.91} & 81.01 & \multicolumn{1}{c|}{82.04} & 84.42 & \multicolumn{1}{c|}{\textbf{8.57}} & \textbf{0.00} & \multicolumn{1}{c|}{\textbf{27.00}} & \textbf{0.00} & \multicolumn{1}{c|}{49.63} & 41.360 \\ \hline
\textbf{SwAV} & \multicolumn{1}{c|}{82.32} & 77.84 & \multicolumn{1}{c|}{80.61} & 84.59 & \multicolumn{1}{c|}{\textbf{9.75}} & \textbf{0.00} & \multicolumn{1}{c|}{\textbf{12.66}} & \textbf{0.00} & \multicolumn{1}{c|}{46.33} & 40.608 \\ \hline
\textbf{MoCo v2} & \multicolumn{1}{c|}{82.56} & 82.27 & \multicolumn{1}{c|}{82.49} & 83.60 & \multicolumn{1}{c|}{\textbf{23.20}} & \textbf{1.40} & \multicolumn{1}{c|}{\textbf{21.33}} & \textbf{0.00} & \multicolumn{1}{c|}{52.39} & 41.822 \\ \hline
\end{tabular}
}
\caption{Results of the experiments with one SSL task with the training and validation data.}
\label{training}
\end{table}

\subsection{Performance results for curricular SSL pretrained models}\label{subsection:results}

\begin{table}[t!]
\centering
\begin{tabular}{clcc}
\hline
\textbf{Curriculum} & \textbf{Pretraining} & \textbf{Validation Acc (\%)} & \textbf{AIL (\%)} \\ \hline
- & \textbf{ImageNet} & 82.75 & \textbf{38.16} \\
- & \textbf{Scratch} & \textbf{83.69} & 37.30 \\ \hline
- & \textbf{Rel-Loc} & 83.62 & 36.33 \\
- & \textbf{MoCo v2} & 83.89 & 37.40 \\
- & \textbf{Swav} & 83.97 & 42.82 \\
- & \textbf{Rotation} & \textbf{84.72} & \textbf{45.95} \\ \hline
\checkmark & \textbf{MoCo v2 + Rotation} & 84.77 & \textbf{47.92} \\
 & \textbf{MoCo v2 + Rel-Loc} & \textbf{85.59} & 39.57 \\
\checkmark & \textbf{MoCo v2 + SwAV} & 83.67 & 41.79 \\ \hdashline
 & \textbf{Rotation + MoCo v2} & 76.08 & 31.87 \\
 & \textbf{Rotation + Rel-Loc} & 84.33 & \textbf{44.48} \\
 & \textbf{Rotation + SwAV} & \textbf{84.81} & 41.46 \\\hdashline
\checkmark & \textbf{Rel-Loc + Rotation} & 84.54 & {\color{blue}\textbf{48.63}} \\
\checkmark & \textbf{Rel-Loc + MoCo v2} & 82.79 & 39.41 \\
\checkmark & \textbf{Rel-Loc + SwAV} & \textbf{85.27} & 46.26 \\\hdashline
\checkmark & \textbf{SwAV + Rotation} & 83.89 & \textbf{47.16} \\
 & \textbf{SwAV + Rel-Loc} & \textbf{84.92} & 43.05 \\
 & \textbf{SwAV + MoCo v2} & 82.37 & 36.51 \\ \hline
 & \textbf{MoCo v2 + Rotation + Rel-Loc} & \textbf{85.28} & \textbf{38.82} \\
 & \textbf{MoCo v2 + Rotation + SwAV} & 84.80 & 30.69 \\\hdashline
 & \textbf{MoCo v2 + Rel-Loc + Rotation} & 84.19 & 38.51 \\
 & \textbf{MoCo v2 + Rel-Loc + SwAV} & \textbf{85.49} & \textbf{46.30 } \\\hdashline
\checkmark & \textbf{MoCo v2 + SwAV + Rotation} & {\color{blue}\textbf{85.67}} & 40.19 \\
 & \textbf{MoCo v2 + SwAV + Rel-Loc} & 83.74 & \textbf{40.89} \\ \hline
\end{tabular}
\caption{Balanced accuracies and AIL scores for the curricular SSL-task pretraining configurations. Sequential orderings for SSL-tasks read left to right. The curriculum column indicates whether a SSL-task combination follows a curriculum ordering. Results in bold refer to the highest score of each block, while results in blue are the highest scores overall.}
\label{results}
\end{table}

The balanced accuracy performances for models trained with different curricular SSL pretraining configurations on two classes: \textit{Negative} and \textit{Typical}, together with the two baselines\textemdash ImageNet, and Scratch, can be found in Table~\ref{results}. We have tested single-task pretraining, all possible combinations of two-task pretraining, and a subset of combinations of three-task pretraining. The selection of the subset of three-task pretraining combinations is defined according to the results of two-task ones: all two-task combinations ending with MoCo v2 decrease the performance with respect to the single-task alternative, e.g,
Rotation Prediction + MoCo v2 presents an accuracy (76.08\%) lower than Rotation Prediction (84.72\%), and even the baselines. Conversely, MoCo v2 + Relative-Location (which starts with MoCo v2) results in the highest accuracy among all two-task combinations (85.59\%). Therefore, we evaluated all possible three-task combinations with MoCo v2 as the first pretraining step. 

% However, further research needs to be done to study this problem. 
Performances in Table \ref{results} indicate that there is a significant gap in performance between the baseline models: training from Scratch (83.69\%) achieves a better performance than supervised pretraining on ImageNet (82.75\%). This may be caused by a large domain gap between a natural imagining dataset (ImageNet) and the target CXR dataset. Among single SSL-tasks pretraining task, Rotation Prediciton  achieves the best performance (84.72\%), outperforming both baselines. Performance is further improved for the best combination of two SSL-tasks (MoCo v2 + Relative Location; 85.59\%) and three (MoCo v2 + SwAV + Rotation Prediction; 85.67\%). All together, these performances experimentally validate our hypotheses: (i) SSL pretraining benefits performance and (ii) some sequential combinations of SSL tasks further improve performances. However, the performance gain is smaller as additional tasks are included, suggesting that performance may be saturating.

In~\cite{sirotkin2021improved}, curricular SSL orderings are defined based on the accuracy achieved by single-SSL task pretraining plus downstream task fine-tuning. The \textit{curriculum} order corresponds to a purely incremental accuracy order. This curriculum premise is fullfilled for the best three-task combination. For two SSL-tasks, the second best option (Relative Location + SwAV: 85.27\%) also follows a curriculum order, indicating that arranging the SSL tasks in a curriculum order yields a performance competitive with top-performance combinations.

\subsection{Model attention results: AIL score} \label{subsection:ail}

% \todo[inline,color=green]{pcl: Specify that reported AIL scores are obtaned an average value of the validation split}

\begin{figure}
    \centering
    \includegraphics[width=\linewidth]{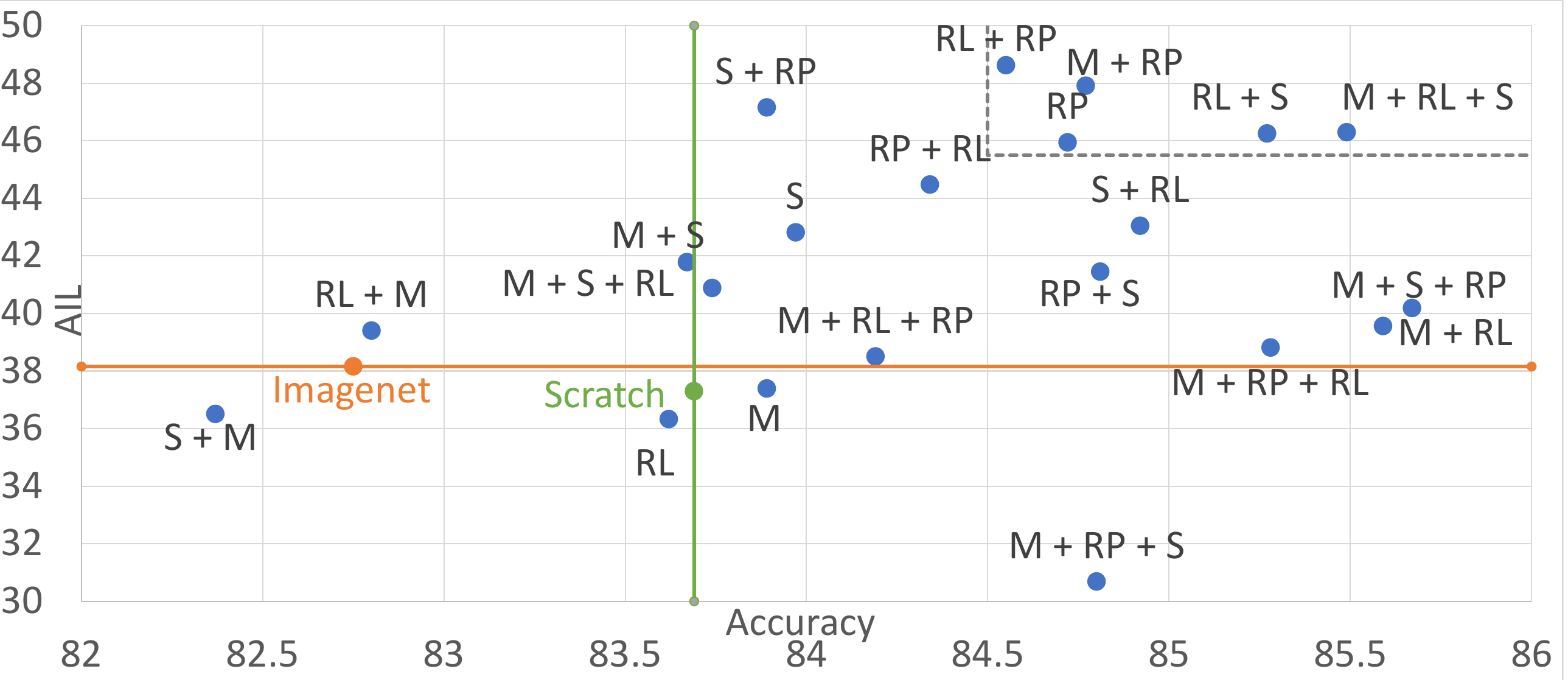}
    \caption{Relation between accuracy and AIL scores for the evaluated meodels. The green line indicates the best baseline accuracy (Scratch), while the orange indicates the highest baseline AIL score (Imagenet). The labels refer to: MoCo v2 (M), SwAV (S), Relative Location (RL) and Rotation Prediction (RP). For the sake of visualization, the RP+M is removed from the graph (Acc. 76.08\%).}
    \label{ail-acc}
\end{figure}

This section provides a comparison on the degree of attention in the lung region, among the models evaluated in Section \ref{subsection:results}, using the average AIL score described in Section \ref{subsection:attention}. The results for all the models, averaged for all validation images, are shown in Table~\ref{results}, and Figure~\ref{ail-acc} shows the correlation between the AIL score and accuracy for each model. The results show that 15 out of 21 SSL-pretrained models show an AIL score exceeding that of the ImageNet baseline (38.16\%) and accuracy higher than that of the Scratch baseline (83.69\%).
% \todo[inline,color=yellow]{pcl: this next is not very specific.Indicate in the figure}
The figure also demonstrates that the models with a high AIL score, and thus, focus more in the lung area, have a grater performance. These are the models within the area are bounded in the top right corner of Figure~\ref{ail-acc}, yielding top-accuracy values and a high AIL score. This gives an indication that SSL-pretraining may be beneficial to obtain accurate models that show a higher degree of attention in the lung regions, and thus, are more robust to confounding factors in the training dataset, as described in Section~\ref{section:sound}. 

% Contrary to the accuracy results, pretraining on ImageNet (38.16\%) gives a better AIL score than training from Scratch (37.30\%). 
Furthermore, there is evidence that multiple-task pretraining increases AIL score, along with accuracy. Regarding single SSL pretraining, the higher AIL score corresponds to Rotation Prediction (45.95\%) which also shows the higher single-task accuracy. For combinations of two SSL tasks, the higher AIL score corresponds to Relative Location + Rotation Prediction (48.63\%)\textemdash the best combination of models among the studied ones. Interestingly, the highest AIL score in each block of two SSL-tasks combinations ends with Rotation Prediction, which also yields the highest AIL score individually. Finally the best combination of 3 SSL tasks is MoCo v2 + Relative Location + SwAV (46.30\%).

\paragraph{Experimental verification on the existence of confounding factors}
Finally, leveraging the lung segmentation strategy, we have performed an additional analysis that corroborates the existence of confounding factors out of the lung regions. Specifically, we have trained the Scratch and ImageNet baselines using inversely-segmented images, i.e, removing the lung regions from the image (an example can be found in Figure \ref{masks}) achieving a 71.83\% and a 75.31\% of balanced accuracy respectively. These results are significantly over the 50\% threshold that corresponds to a random guess. This result confirms that  data outside the lung regions can act as a spurious shortcut (which can be dataset dependent) for the pneumonia classification problem. Thus, models with a higher degree of attention in the lung regions (such as SSL-pretrained ones) may embody models with better generalization capabilities.

\section{Conclusions} \label{section:conclusions}
In this work, we evaluated the performance gain of curricular SSL training scheme for pneumonia  classification in CXR images of Covid-19 patients. We show that a combination of SSL tasks can outperform pretraining on ImageNet, or training directly with the target data. With our best configuration, MoCo v2 + SwAV + Relative Location, we achieve a \textbf{+1.98\%} accuracy increase over the baselines. The results provide evidence that additional SSL tasks can increase the performance of the model compared to pretraining with only one SSL task. Finally, an SSL task ordering based on single-SSL pretraining accuracies (curriculum order) results which are the best (or close) to all evaluated combinations.

Moreover, we show that curricular SSL pretraining scheme helps to build models that are more robust to the possible existence of confounding factors outside the lung region, which may avoid overfitting to the training dataset and to increase the generalization capabilities of the model. This is assessed by means of a novel attention sore (AIL) that captures the level of attention of a model in an image region\textemdash lungs in this case. The importance behind creating robust models is to avoid overfitting to the training dataset and to increase the generalization capabilities of the model in other datasets.

%&Addressing this issue, we show that our models better focus their attention, according to a novel metric, AIL. 

% We confirm that these external factors were present in the studied dataset by training our models with inverted segmented images.

% \todo[inline, color=yellow]{I would not put this last paragraph. kis: me neither}
% We also confirmed that the \textit{Indeterminate} and \textit{Atypical} classes from the SIIM-FISABIO-RSNA COVID-19 Detection dataset are troublesome due to inconsistencies in the labels and high intra-class variance along with low amount of data respectively. We removed these classes from our work to avoid noise in the results of combining SSL tasks.

\bibliography{egbib}
\end{document}